\documentclass[aip,amsmath,amssymb,reprint,cha]{revtex4-1}

\usepackage{graphicx}
\usepackage{dcolumn}
\usepackage{bm}
\usepackage{physics}
\usepackage[utf8]{inputenc}
\usepackage[T1]{fontenc}
\usepackage{mathptmx}
\usepackage{etoolbox}
\usepackage[hidelinks]{hyperref}
\makeatletter
\def\@email#1#2{%
 \endgroup
 \patchcmd{\titleblock@produce}
  {\frontmatter@RRAPformat}
  {\frontmatter@RRAPformat{\produce@RRAP{*#1\href{mailto:#2}{#2}}}\frontmatter@RRAPformat}
  {}{}
}%

\renewcommand*\email[1][Authors to whom correspondence must be addressed: ]{\begingroup\sanitize@url\@email{#1}}%

\makeatother
\begin{document}

\preprint{AIP/123-QED}

\title[Photoluminescent cooling with incoherent light]{Photoluminescent cooling with incoherent light}
\author{Sushrut Ghonge}
\affiliation{Department of Physics and Astronomy, University of Notre Dame, Notre Dame, Indiana USA.}
\email{sghonge@nd.edu, bjanko@nd.edu}
 
\author{Masaru Kuno}
\affiliation{Department of Chemistry and Biochemistry, University of Notre Dame, Notre Dame, Indiana USA.}
\affiliation{Department of Physics and Astronomy, University of Notre Dame, Notre Dame, Indiana USA.}

\author{Boldizs\'{a}r Jank\'{o}}
\affiliation{Department of Physics and Astronomy, University of Notre Dame, Notre Dame, Indiana USA.}

\date{\today}

\begin{abstract}
Optical refrigeration using anti-Stokes photoluminescence is now well established, especially for rare-earth-doped solids where cooling to cryogenic temperatures has recently been achieved. The cooling efficiency of optical refrigeration is constrained by the requirement that the increase in entropy of the photon field must be greater than the decrease in entropy of the sample. Laser radiation has been used in all demonstrated cases of optical refrigeration with the intention of minimizing the entropy of the absorbed photons. Here, we show that as long as the incident radiation is unidirectional, the loss of coherence does not significantly affect the cooling efficiency. Using a general formulation of radiation entropy as the von Neumann entropy of the photon field, we show how the cooling efficiency depends on the properties of the light source such as wavelength, coherence, and directionality. Our results suggest that the laws of thermodynamics permit optical cooling of materials with incoherent sources such as light emitting diodes and filtered sunlight almost as efficiently as with lasers. Our findings have significant and immediate implications for design of compact, all-solid-state devices cooled via optical refrigeration. 
\end{abstract}

\maketitle

\section{\label{sec:level1}Introduction}

Laser cooling of atomic gases has proven to be a successful method of achieving ultralow temperatures \cite{cohen1998nobel,chu1998nobel,phillips1998nobel} and symmetry broken condensed phases \cite{anderson1995observation} such as Bose-Einstein condensates. While cooling semiconductors remains elusive,\cite{zhang2013laser,morozov2019can,zhang2019progress} rare-earth-doped solids have been cooled from room temperature to $<$ 100 K using laser cooling. \cite{epstein1995observation,seletskiy2010laser,melgaard2013optical,melgaard2016solid} In this Letter, we theoretically study if coherent light sources like lasers are necessary for optical cooling, and how much cooling, if any, can be achieved with incoherent sources like Light Emitting Diodes (LEDs) and filtered thermal radiation.

Anti-Stokes photoluminescence (ASPL), a process where a material emits higher energy photons than it absorbed, underlies laser cooling. The energy difference between absorbed and emitted photons, known as detuning, is supplied by heat from the sample. Therefore, ASPL can be used to cool a material. In laser cooling, photons act as the ``working fluid" of the optical refrigerator, converting heat from the material into emitted photons. Transitions between the ${}^2F_{5/2} \text{ and } {}^2F_{7/2}$ manifolds of Yb$^{3+}$ have been exploited to laser cool Yb:YLiF$_{4}$ from room temperature to $<$ 100 K using ASPL. \cite{epstein1995observation,seletskiy2010laser,melgaard2013optical,melgaard2016solid}

The second law of thermodynamics requires that the increase in entropy of photons must be more than the decrease in entropy of the sample being cooled, thus imposing constraints on the cooling efficiency. This fact has been used to derive thermodynamic constraints on cooling efficiency by various authors. \cite{mungan1997laser,mungan1999laser,mungan2005radiation,mungan2009thermodynamics,ruan2007entropy,ruan2007advances} Their calculations rely on the fact that the photoluminescent emission has a larger bandwidth and is less unidirectional than the incident radiation. However, all such calculations have used the classical equilibrium formulation of radiation entropy, which is not applicable to nonclassical light sources or nonequilibrium states of the photon field. Expressing the radiation entropy as the von Neumann entropy of the photon field enables us to study classical and non-classical light sources.

Recent experiments on optical cooling by controlling the chemical potential of photons \cite{zhu2019near,buddhiraju2020photonic,schirber2020emitting} and electroluminescence \cite{sadi2020thermophotonic,chen2017high,xiao2018electroluminescent} as well as theoretical efforts addressing optomechanical cooling \cite{mari2012cooling} and electron pumping \cite{cleuren2012cooling} led us to question whether coherence is necessary in ASPL-based optical cooling. In this Letter, we show that so long as the radiation remains unidirectional, loss of coherence does not significantly affect the cooling efficiency. This means that even filtered sunlight can optically cool materials almost as efficiently as lasers. Low intensity unidirectional sources yield a greater cooling efficiency than divergent and/or high intensity sources with the same absorbed power.

\section{Entropy of photon fields}

We model the sample as a collection of photoluminescent emitters with three or more energy levels. In the ASPL process, an incident photon excites the emitter from the ground state to an intermediate state. The emitter then either relaxes to the ground state, or gains thermal energy to reach the excited state before relaxing to the ground state, as shown in \textbf{Figure 1 (a)}. In rare-earth ions, the intermediate state is created by crystal field splitting of ${4}^2F$ states, while energy gain in ASPL is attributed to multiple-phonon absorption. \cite{seletskiy2016laser,zhang2013laser,auzel1976multiphonon,auzel2004upconversion,zhang2024resonant} The thermodynamic constraints derived in this Letter depend only on photon field states and are otherwise independent of the identity of electronic states and the microscopic mechanism of ASPL.

The energy difference between the intermediate and excited states is called detuning, and is denoted by $\Delta E$. The photoluminescence quantum yield $\Phi$ is defined as the ratio of the number of emitted photons to the number of absorbed photons. $\Phi$ is therefore the ratio of the radiative relaxation rate to the total relaxation rate. $\Phi < 1$ indicates the presence of non-radiative relaxation in the sample, an irreversible process that produces heat and hence entropy. The ASPL efficiency $\eta_{\text{ASPL}}$ is defined as the fraction of absorbed photons emitted as high frequency photons. The cooling efficiency $\eta_{\text{c}}$ is the ratio of the cooling power to the absorbed power.

If $S_i$, $S_a$, and $S_u$ are the entropies of the absorbed, anti-Stokes shifted, and unshifted photons (that is, photons emitted at the same wavelength as the incident photons), respectively, and $\bar{Q}$ is the heat removed from the sample, then the second law of thermodynamics states that
\begin{align}\label{law2}
    S_a + S_u - S_i - \bar{Q}/k_BT &\geq 0.
\end{align}

Since we are interested in the steady state cooling efficiency, all entropy and heat changes can be calculated for an arbitrary time interval, which we choose to be the ASPL lifetime $\tau$. In steady state, the sample reaches a balance between heat inflow from the environment and heat removal with ASPL, assuming that the environment remains at 300 K. With $\nu_i$ and $\nu_a$ being the mean frequencies of incident and anti-Stokes shifted photons, respectively, and $\bar{n}$ being the mean number of incident photons in one ASPL lifetime, 
\begin{equation}\label{heatrem}
    \bar{Q} = h\bar{n} [\eta_{\text{ASPL}}\Phi  (\nu_a - \nu_i) - (1-\Phi) \nu_i].
\end{equation}

The heat obtained from this equation is substituted into the the second law (Eq. \ref{law2}) to obtain the maximum possible value of $\eta_{\text{ASPL}}$. $\bar{n}$ is related to the absorbed power $P$, the ASPL lifetime $\tau$, and $\nu_i$ as $\bar{n} = P\tau/h\nu_i$. The associated cooling efficiency $(\eta_{\text{c}}=\bar{Q}/\bar{n}h\nu_i)$ is 
\begin{equation}\label{coolingeff}
    \eta_{\text{c}} = \Phi\eta_{\text{ASPL}}\Delta E/h\nu_{i} - (1-\Phi).
\end{equation}

The entropy of photons ($S_i$, $S_a$ and $S_u$) from any source is the von Neumann entropy of the photon field, $S= -k_B \Tr{\hat{\rho}\log(\hat{\rho})}$, where $\hat{\rho}$ is the state of the photon field. This formulation differs from the classical formulations of Landau, Mungan, and others \cite{landau1946thermodynamics,mungan2009thermodynamics,ruan2007entropy,mungan2005radiation} in that the radiation needs to be coherent to have zero entropy. Fock states (occuring in optical cavities), coherent states, and their coherent superpositions are pure states. Thus, their von Neumann entropy is zero.

A single mode laser operating far above the threshold is represented in the Sudarshan-Glauber P-representation \cite{sudarshan1963equivalence,glauber1963coherent,perina1985coherence} by $\mathcal{P}(\alpha) = \frac{1}{2\pi\abs{\alpha}}\delta (\abs{\alpha} - \abs{\beta})$, \cite{perina1991quantum,mandel1995optical,scully2019laser} where $\mathcal{P}(\alpha)$ is a quasiprobability distribution over coherent states with eigenvalues $\alpha$ and amplitude $\abs{\beta}$.
This is equivalent to a density operator $\hat{\rho} = \sum_n \ket{n}\bra{n} \frac{e^{-n}}{n!}\abs{\beta}^{2n}$, corresponding to a Poisson distribution over number eigenstates with $\bar{n} = \abs{\beta}^2$. Using the entropy of the Poisson distribution \cite{evans1988entropy}, we find the von Neumann entropy of each mode of a laser ($S_L$) to be

\begin{equation}\label{laserent}
    S_L = k_B\left\{\bar{n}[1 - \log(\bar{n})] + e^{-\bar{n}} \sum_{m=0}^{\infty} \frac{\bar{n}^m \log(m!)}{m!}  \right\}.
\end{equation}

The entropy of a single mode of radiation at thermal equilibrium $S_B$ is the entropy function for bosons in a single mode \cite{bose1924plancks,toda2012statistical}, $S_{\text{B}}(\bar{n}) = k_B [(\bar{n}+1)\log(\bar{n}+1) - \bar{n} \log(\bar{n}) ].$ The same expression is also valid for any non-equilibrium distribution for which $P(n) \propto q^n$ for any real number $q$. \cite{landsberg1980thermodynamic}

For a partially coherent source, the von Neumann entropy must be evaluated numerically from the density operator. However, one can also derive analytical expressions by assuming that different photon modes are independent of each other. Under this assumption, the total entropy is the sum of the entropies of all the modes, weighted by their occupation numbers. Beam angular divergences and cross sectional areas are assumed in the absence of any lenses or mirrors. This is because they change the angular divergence of beams, thus changing the occupations of different modes of the photon field. This is further demonstrated by the fact that when a parallel beam passes through a convex lens, the photon's wavevector $\vec{k}$ is completely determined by position, making the assumption of uncorrelated modes invalid. Even then, ideal lenses and mirrors are passive devices that do not change the entropy of the photon field, and hence do not affect the cooling efficiency.

\begin{figure*}
    \centering
    \includegraphics[height=0.3\linewidth]{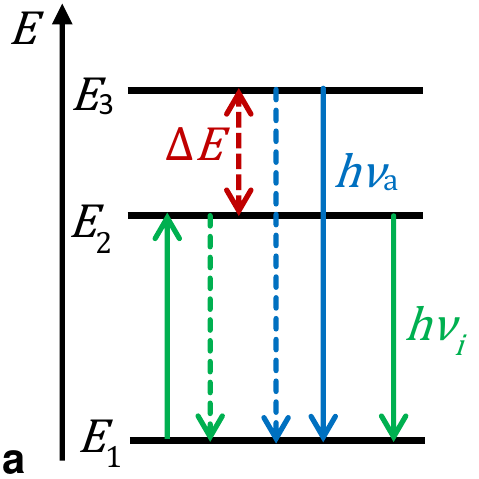}
    \hspace{0.05in}
    \includegraphics[height=0.3\linewidth]{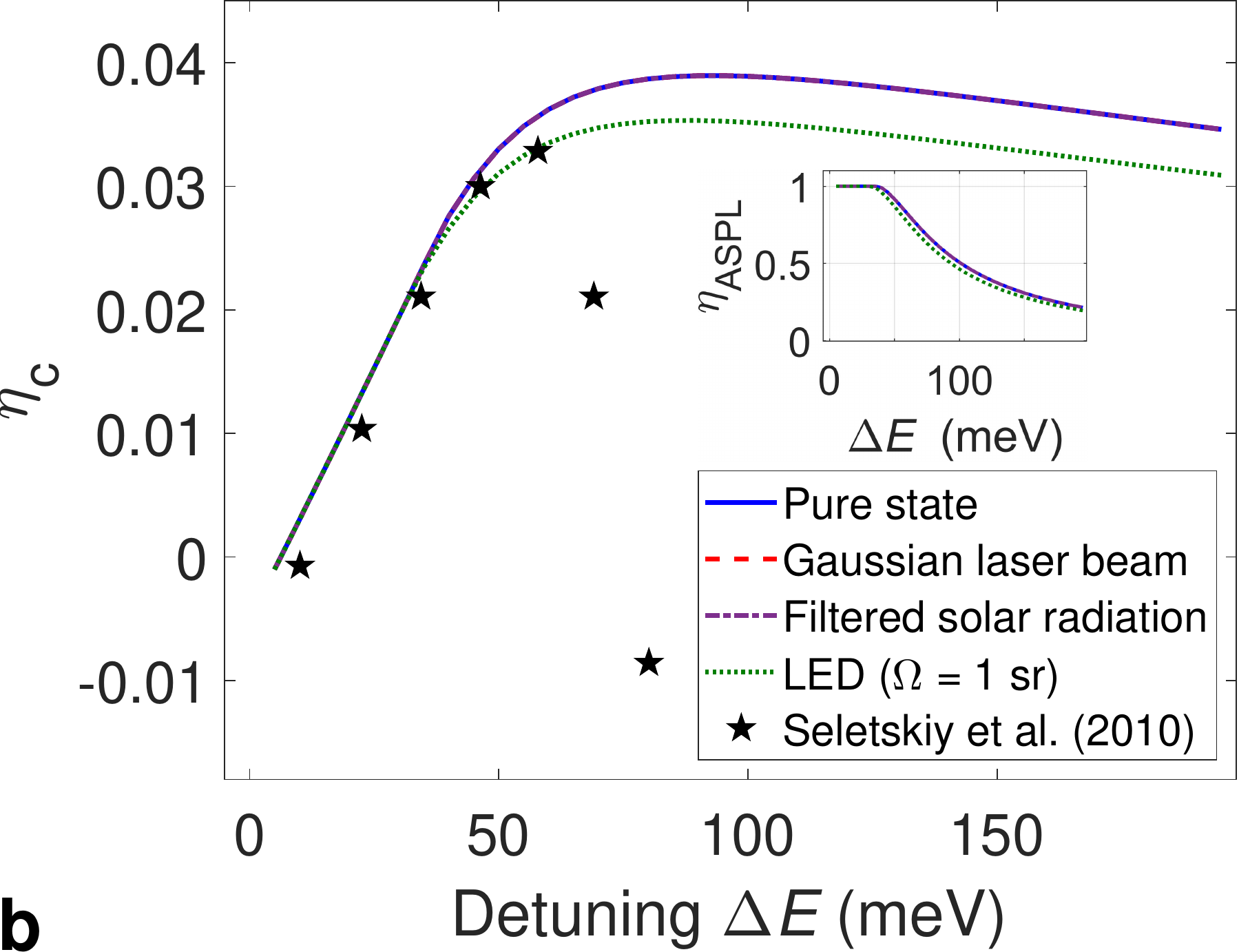}
    \hspace{0.05in}
    \includegraphics[height=0.3\linewidth]{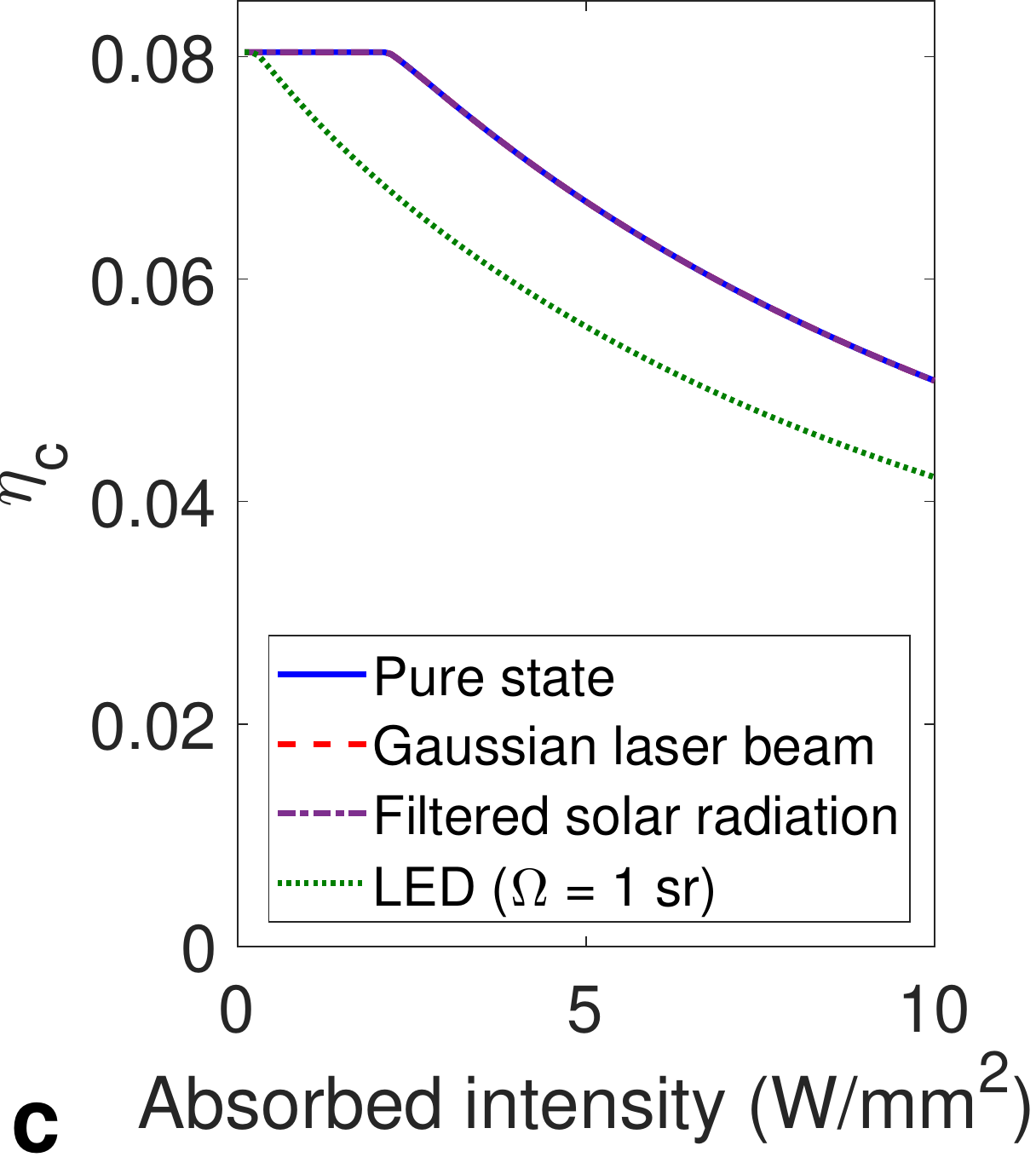}
    \caption{(a) Schematic of ASPL-based cooling. (Dashed) solid arrows show (non-)radiative excitation and relaxation processes. (b) Comparison of the maximum possible cooling efficiency (lines) and the cooling efficiency observed by Seletskiy \textit{et al.} \cite{seletskiy2010laser} (pentagrams) for different detunings and sources of incident radiation. All parameters are as reported by Seletskiy \textit{et al.}, \cite{seletskiy2010laser} which corresponds to an intensity of $~50$ W/mm$^2$. (c) Dependence of maximum cooling efficiency on absorbed intensity for $\Delta E$ = 100 meV; all other parameters are the same as (b). The inset in panel (b) shows the maximum ASPL efficiency for the same parameters.}
    \label{fig:detuning_source}
    \hrulefill
\end{figure*}

The density of states of photons with an absolute wavevector $k$ and in a solid angle $\text{d}\Omega$ is $g(k,\Omega) = 2k^2 \text{d}k \text{d}\Omega$, where $k = 2\pi\nu/c$. In general, the total number of states of the photon field in a volume, $V$, is the integral of the density of states, which for isotropic and nearly monochromatic radiation gives $ N(k) = 8\pi k^2 (\Delta k) V$. $V$ is the product of the beam cross section area, $A$, and the length, $c\tau$, associated with the ASPL lifetime ($\tau$), where $c$ is the speed of light. Therefore, the entropy of isotropic, filtered thermal radiation with a cross section area $A$ and mean photon number $\bar{n}$ is $S_T = (8\pi k^2) S_B(\bar{n})  A  (c\tau)(\Delta k).$

The line-broadening of absorption and emission is related to $\tau$ through the Wigner-Weisskopf linewidth-lifetime relation (shown to be valid even in the case of dissipative peak broadening by Stedman \cite{stedman1971validity}) as $(\Delta\nu)\tau=(c\Delta k)(\tau)=1$. This simplifies the expression for $S_T$ to $(8\pi k^2A) S_B(\bar{n})$.  If the photons are confined to a solid angle $\Omega$, $N(k)\text{d}k = 2 \Omega k^2 \text{d}k V$ and $S_T = 2 \Omega k^2 S_B(\bar{n})$.

A Gaussian laser beam in the diffraction limit satisfies $\Delta\theta = 2/kR$, where $\Delta\theta$ is the beam divergence, $R$ is the radius of the beam spot and $k=\abs{\vec{k}}$. Using this, the entropy of a Gaussian laser beam ($S_G$) is,
\begin{equation}
    S_G = 2 k^2 A \sin^2(\Delta\theta) S_L(\bar{n}) = 8\pi S_L(\bar{n}).
\end{equation}
\noindent We substitute these expressions for the entropy of different light sources into Eq. \ref{law2} to obtain $\eta_{\text{ASPL}}$ for various beam divergences and diameters, mean emission frequencies, and sample external quantum efficiencies. For example, if incident photons come from a Gaussian laser, and the emitted photons are incoherent, then $S_i$ is substituted by $S_G$ with the corresponding frequency and intensity, while $S_u$ and $S_a$ are substituted by $S_T$ using the corresponding unshifted and anti-Stokes shifted frequencies. We subsequently obtain $\eta_{\text{c}}$ using Eq. \ref{coolingeff}.

\section{Thermodynamic limits on cooling efficiency}

While our results are applicable to any material, we specifically apply our theory to Yb:LiYF$_4$, which has been laser-cooled to cryogenic temperatures, \cite{seletskiy2010laser,melgaard2013optical,melgaard2016solid} and for which all the parameters necessary for our theoretical modelling have been experimentally determined and reported. We use the mean ASPL emission energy (1.25 eV), absorbed power (3.5 W), beam radius (150 $\mu$m), photoluminescence quantum yield ($\Phi \approx 0.995$), and absorption linewidth (20 meV) reported by Seletskiy \textit{et al.} \cite{seletskiy2010laser}. We further show the dependence of $\eta_{\text{c}}$ on the mean ASPL emission energy, beam radius, solid angle subtended by the source, and $\Phi$.

\begin{figure}
    \centering
    \includegraphics[height=0.55\linewidth]{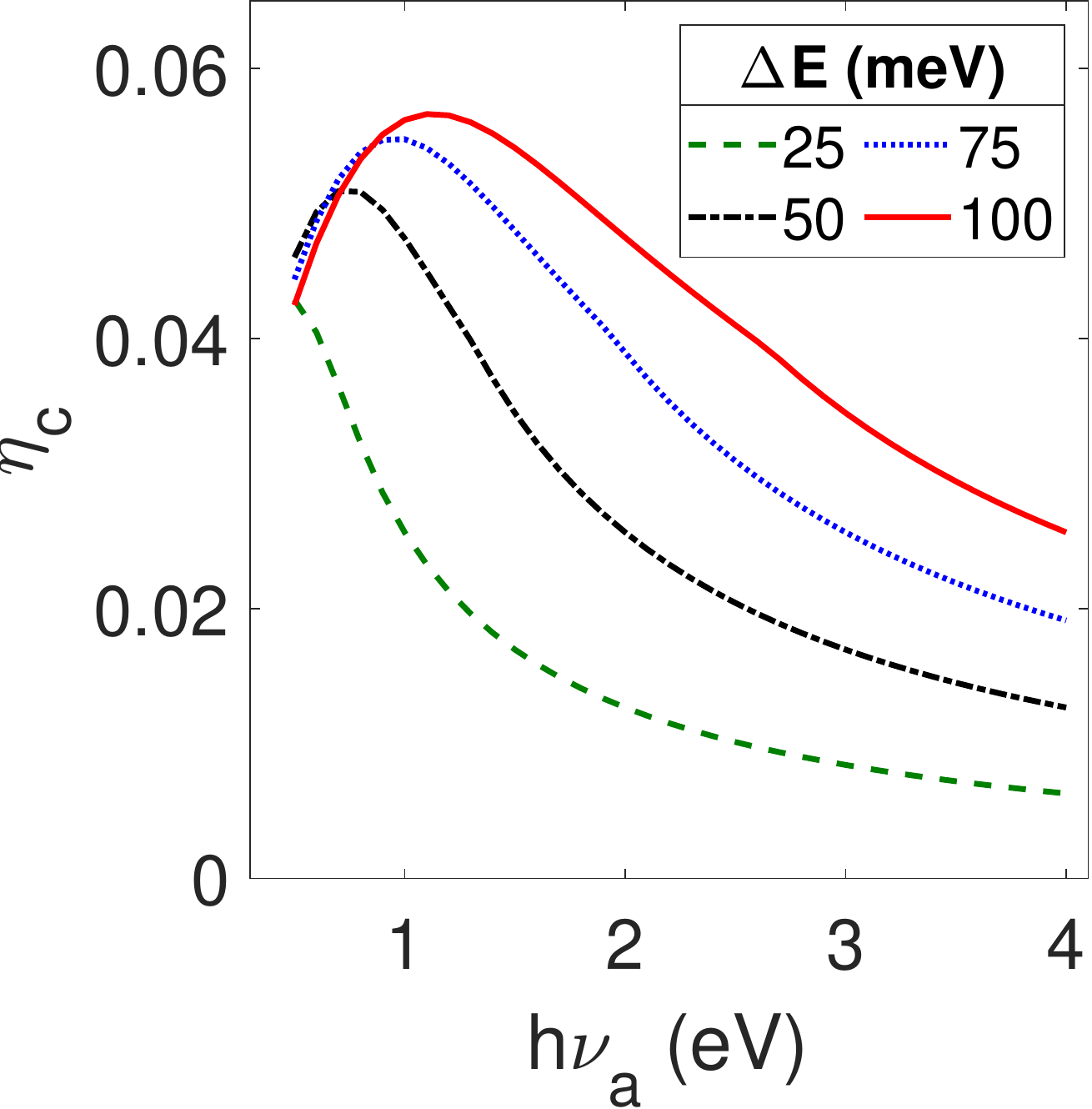}
    \caption{Dependence of $\eta_{\text{c}}$ on the ASPL energy $h\nu_a$ and detuning for a laser source. The photon flux is 6.84 $\times 10^{4}$ photons/$\mu$m$^2$ns (corresponding to an intensity of 13.1 W/mm$^2$ when the absorbed photon energy is 1.2 eV, and an absorbed power of 0.93 W for a beam radius of 150 $\mu$m), and $\Phi = 1$.}
    \label{fig:bg_source}
    \hrulefill
\end{figure}

\begin{figure}
    \centering
    \includegraphics[height=0.6\linewidth]{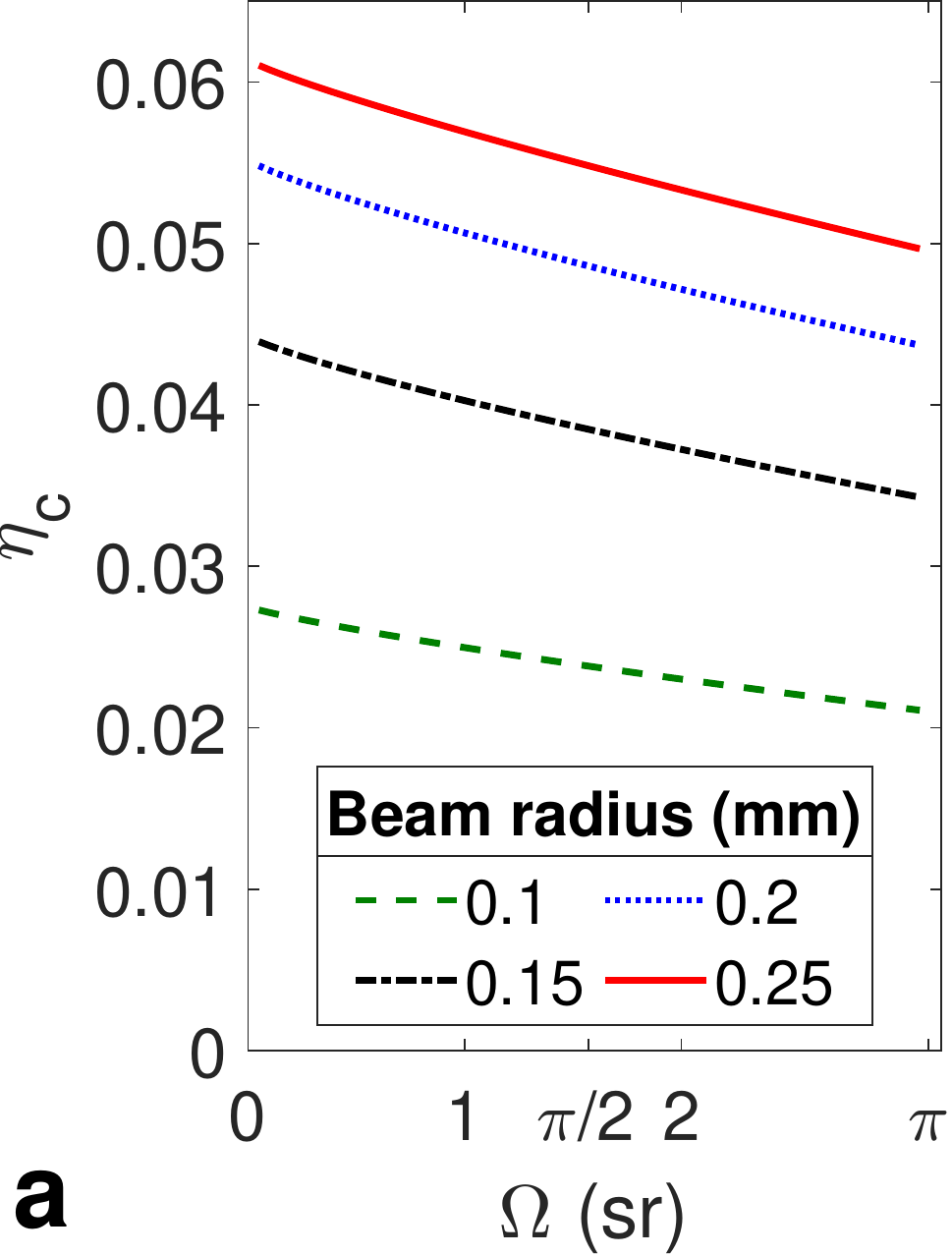}
    \hspace{.1in}
    \includegraphics[height=0.6\linewidth]{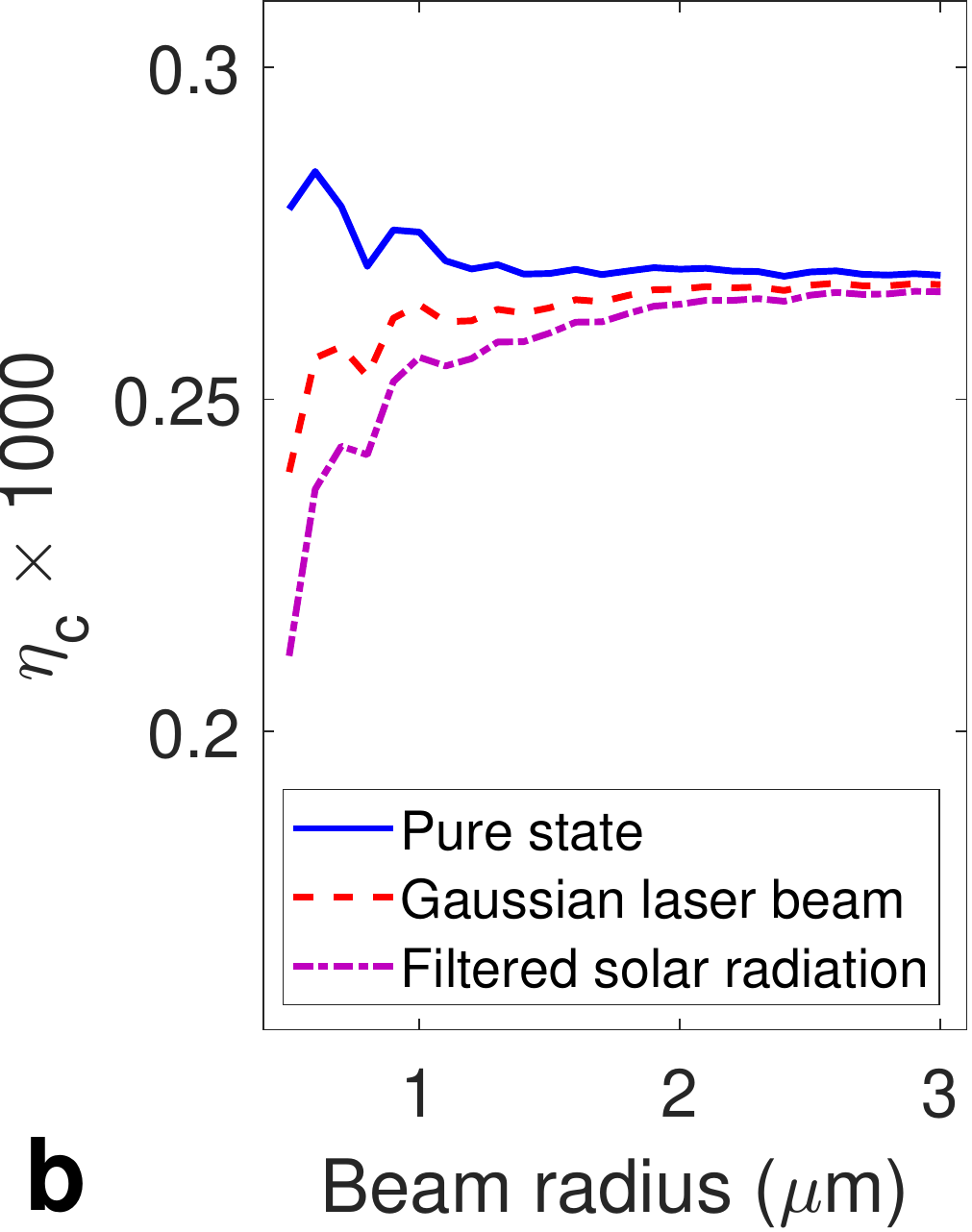}
    \caption{(a) Dependence of $\eta_{\text{c}}$ on the solid angle $\Omega$ subtended by the source on the sample for different beam radii. The absorbed power is 2.7 W, corresponding to a source intensity of 13.9 W/mm$^2$ when the beam radius is 0.25 mm. (b) Dependence of $\eta_{\text{c}}$ on beam radius close to the wavelength of absorbed photons (1.055 $\mu$m at 1.175 eV). Intensity is fixed and absorbed power varies from 6.7 to 241 mW with beam radius. In both panels, the mean ASPL emission energy is 1.25 eV, $\Delta E = 75$ meV, and $\Phi = 1$.}
    \label{fig:detuning_angle}
    \hrulefill
\end{figure}

\textbf{Figure \ref{fig:detuning_source} (b)} shows $\eta_{\text{c}}$ for different sources and frequencies of incident photons and fixed absorbed power. Evident is that $\eta_{\text{c}}$, for a pure state of the photon field (\textit{e.g.,} produced by a perfectly coherent source), a Gaussian laser beam, or filtered far-field thermal radiation, are nearly equal. This is because the contribution of decoherence to the total photon entropy is much less than that of increase in the frequency and solid angle of emission. 

$\eta_{\text{c}}$ decreases with increasing absorbed intensity, as shown in \textbf{Fig. 1 (c)}. This means that lower intensity beams can lead to a greater cooling efficiency for the same cooling power. The solar spectral intensity at 1000 nm at sea level is $0.5\ \text{W}/(\text{m}^2\text{meV})$  \cite{mecherikunnel1980spectral}, and the solar angular diameter is $9.3 \times 10^{-3}$ rad. Parabolic reflectors with a 0.6 m diameter/0.15 m effective focal length (making an image of the sun with 1.4 mm diameter), and band pass filters with a peak wavelength of $980 \pm 2$ nm and a bandwidth of $10 \pm 2$ nm are readily available commercially. Using these, one can produce filtered solar photons with $1.265 \pm 0.006$ eV energy and $\approx 1.2$ W/mm$^2$ intensity. The corresponding absorbed power is 1.85 W with a cooling power of 150 mW.

\textbf{Figure \ref{fig:detuning_source} (b)} also shows that our theoretical predictions are consistent with the cooling efficiency observed by Seletskiy \textit{et al.} \cite{seletskiy2010laser} for $\Delta E < 60$ meV. For larger $\Delta E$, the observed cooling efficiency is much less than the maximum theoretically allowed efficiency and $\eta_{\text{c}}$ decreases with increasing $\Delta E$. These observations are equivalently explained by two different lines of reasoning.

First, the inset of \textbf{Fig. \ref{fig:detuning_source} (b)} shows that as the detuning increases, $\eta_{\text{ASPL}}$ is constant for small $\Delta E$. As $\Delta E$ increases, each emitted photon removes more heat from the sample, thus explaining the nearly linear behavior of $\eta_{\text{c}}$ in this regime. For large $\Delta E$, as $\Delta E$ increases, the probability that an electron gains enough energy from heat to transition from the intermediate state to the exited state decreases, as indicated by the decreasing $\eta_{\text{ASPL}}$. Thus, the cooling efficiency peaks and decreases with increasing $\Delta E$. The rate of multiple-phonon absorption, the microscopic mechanism underlying ASPL \cite{seletskiy2016laser,zhang2013laser,auzel1976multiphonon,auzel2004upconversion,zhang2024resonant}, decreases exponentially with $\Delta E$, which explains why the observed $\eta_{\text{c}}$ is less than its maximum possible value for large $\Delta E$.

We arrive at the same conclusion by considering only photon entropy. For the same entropy of absorbed photons, Eqs. \ref{law2} and \ref{heatrem} imply that $\eta_{\text{ASPL}}$ will increase if the entropy of emitted photons increases. If all other parameters are held constant, the photon entropy depends on the frequency through the photon number of states ($S(\nu)\propto N(\nu) \propto \nu^2$). At low $\Delta E$, the unshifted and anti-Stokes shifted photons have nearly the same frequency and hence nearly the same entropy. Thus, both emissions contribute equally to the total entropy of emitted photons, which allows for high $\eta_{\text{ASPL}}$. As $\Delta E$ increases (for fixed $\nu_a$), the frequency and hence the number of modes of the unshifted photons decreases. This leads to a smaller contribution to the emitted photon entropy from the unshifted emission. $\eta_{\text{ASPL}}$ and hence $\eta_{\text{c}}$ therefore decrease.

\textbf{Fig. \ref{fig:bg_source}} shows the dependence of $\eta_{\text{c}}$ on the mean ASPL emission energy for fixed detuning for different sources of incident photons. For any $\Delta E$, there exists an optimal $\nu_a$ that yields a maximum $\eta_{\text{c}}$ just as there is an optimal $\Delta E$ for a given ASPL emission energy.

\begin{figure}
    \centering
    \includegraphics[height=0.58\linewidth]{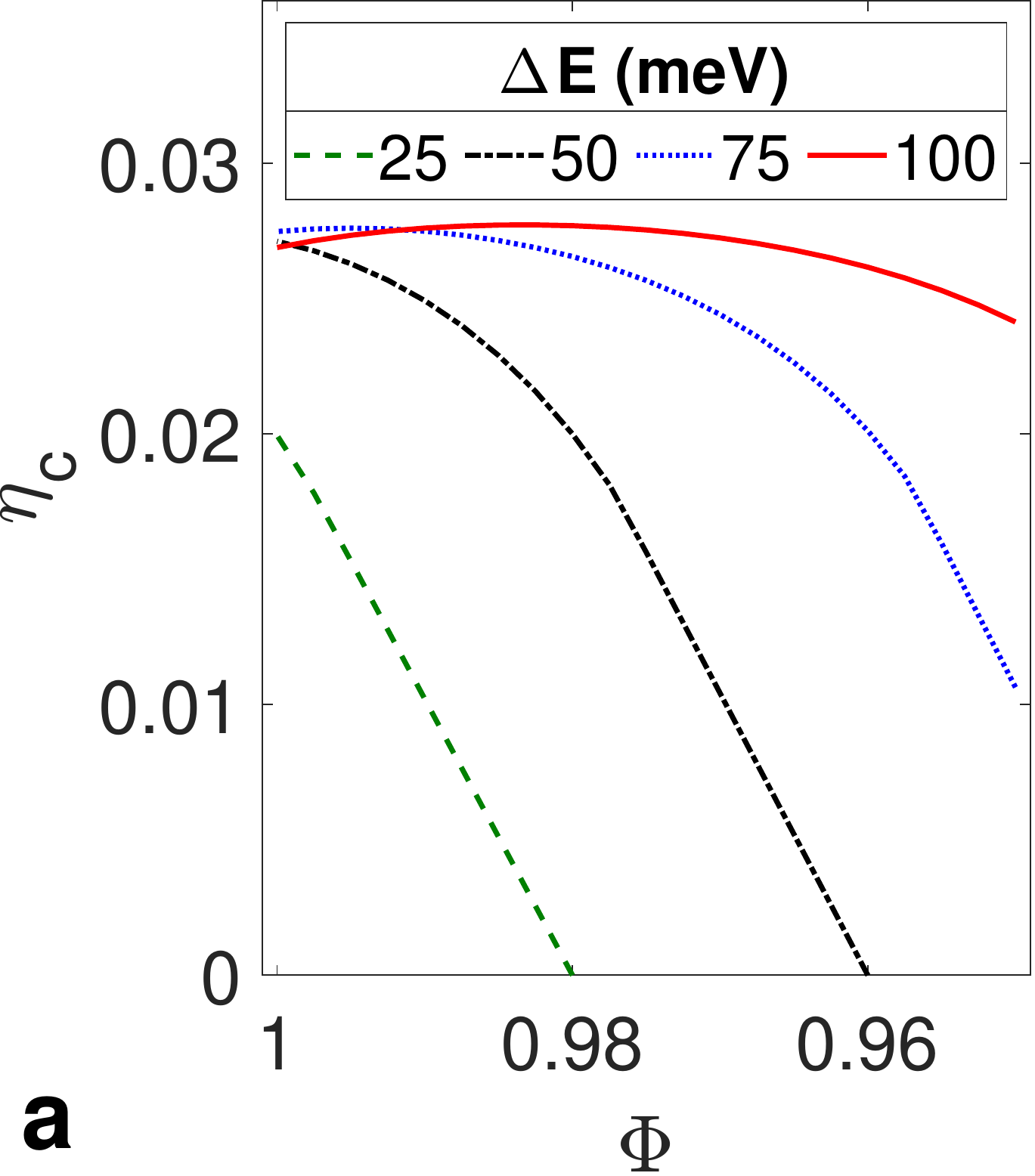}
    \hspace{.1in}
    \includegraphics[height=0.58\linewidth]{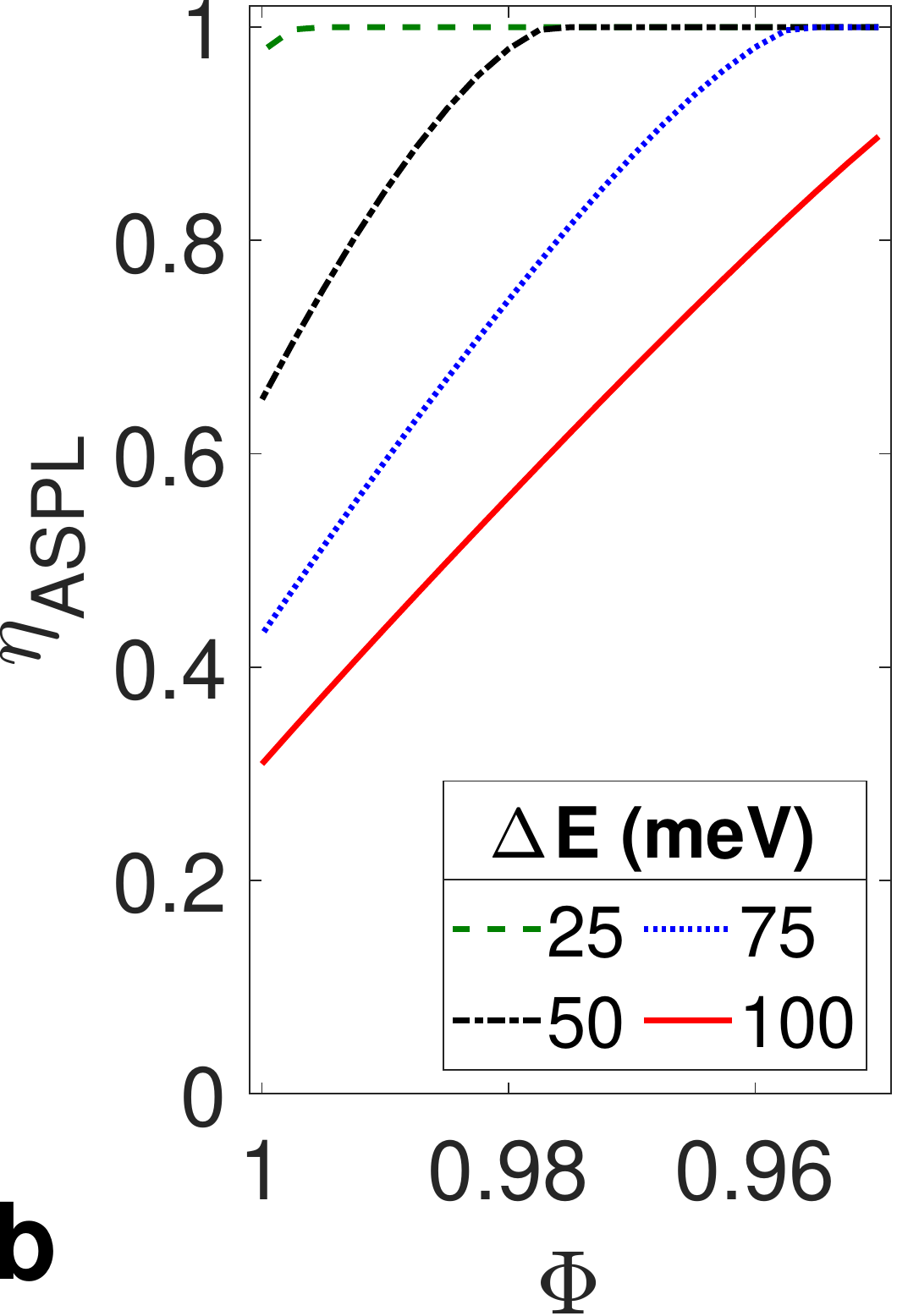}
    \caption{Dependence of (a) $\eta_{\text{c}}$ and (b) $\eta_{\text{ASPL}}$ on the detuning and $\Phi$. The mean ASPL emission energy is 1.25 eV, beam radius is 0.1 mm, and the absorbed power is 2.7 W when the detuning is 75 meV.}
    \label{fig:qy_detuning}
    \hrulefill
\end{figure}

Now we turn to the effect of source unidirectionality and beam cross section area on the cooling efficiency. \textbf{Figure \ref{fig:detuning_angle} (a)} shows that as the beam becomes more divergent, $\eta_{\text{c}}$ decreases. This is because the entropy of the incident photons increases with increasing angular divergence. Note that $\Omega$ in this figure is the solid angle that the source would have subtended in the absence of any lenses or mirrors. \textbf{Fig. \ref{fig:detuning_angle} (a)} also shows that $\eta_{\text{c}}$ increases as beam surface area increases for a fixed absorbed power, as we saw in \textbf{Fig. \ref{fig:detuning_source} (c)}. To our knowledge, the scaling of photon entropy with beam surface area has not hitherto been systematically taken into account in the context of photoluminescence. \footnote{For example, Landau \cite{landau1946thermodynamics} considers radiation passing through a 1 cm$^2$ area in 1 second, while Mungan \textit{et al.} \cite{mungan1997laser,mungan1999laser,mungan2005radiation,mungan2009thermodynamics}, use the entropy per unit area per unit time to define an effective radiation temperature and derive a coefficient of performance for the optical refrigerators similar to the Carnot efficiency.} In the extreme case where beam radius approaches and decreases below the wavelength of incident photons (1.055 $\mu$m at 1.175 eV), cooling becomes relatively inefficient as shown in \textbf{Figure \ref{fig:detuning_angle} (b)}. The contribution of beam divergence to entropy change decreases enough to become comparable to that of decoherence, causing a visible dependence of $\eta_{\text{c}}$ on source coherence, as shown in \textbf{Fig. \ref{fig:detuning_angle} (b)}.

Finally, the dependence of $\eta_{\text{c}}$ on $\Phi$ is shown in \textbf{Fig. \ref{fig:qy_detuning} (a)}, while \textbf{Fig. \ref{fig:qy_detuning} (b)} reveals that samples with $\Phi$ even slightly below unity allow for much higher $\eta_{\text{ASPL}}$. Since $\Delta E$ is much smaller than the energy of incident photons, heat produced in the sample via non-radiative relaxation, following one photon absorption, can compensate for that removed by several ASPL-shifted photons. This means that the entropy change in the sample does not decrease much or can even increase despite $\eta_{\text{ASPL}}$ being near-unity. As the maximum possible $\eta_{\text{ASPL}}$ increases to unity with decreasing $\Phi$, $\eta_{\text{c}}$ remains nearly constant and eventually decreases to negative values due to heating by non-radiative relaxation in the sample [\textbf{Fig. \ref{fig:qy_detuning} (a)}]. For most materials used in optical cooling, the mean emission energy is of the order of an eV or more, while $\Delta E$ can only be of the order of room temperature thermal energy ($\sim 25.8$ meV). These results highlight the importance of having near-unity $\Phi$ in cooling grade samples.

\begin{figure}
    \centering
    \includegraphics[height=0.8\linewidth]{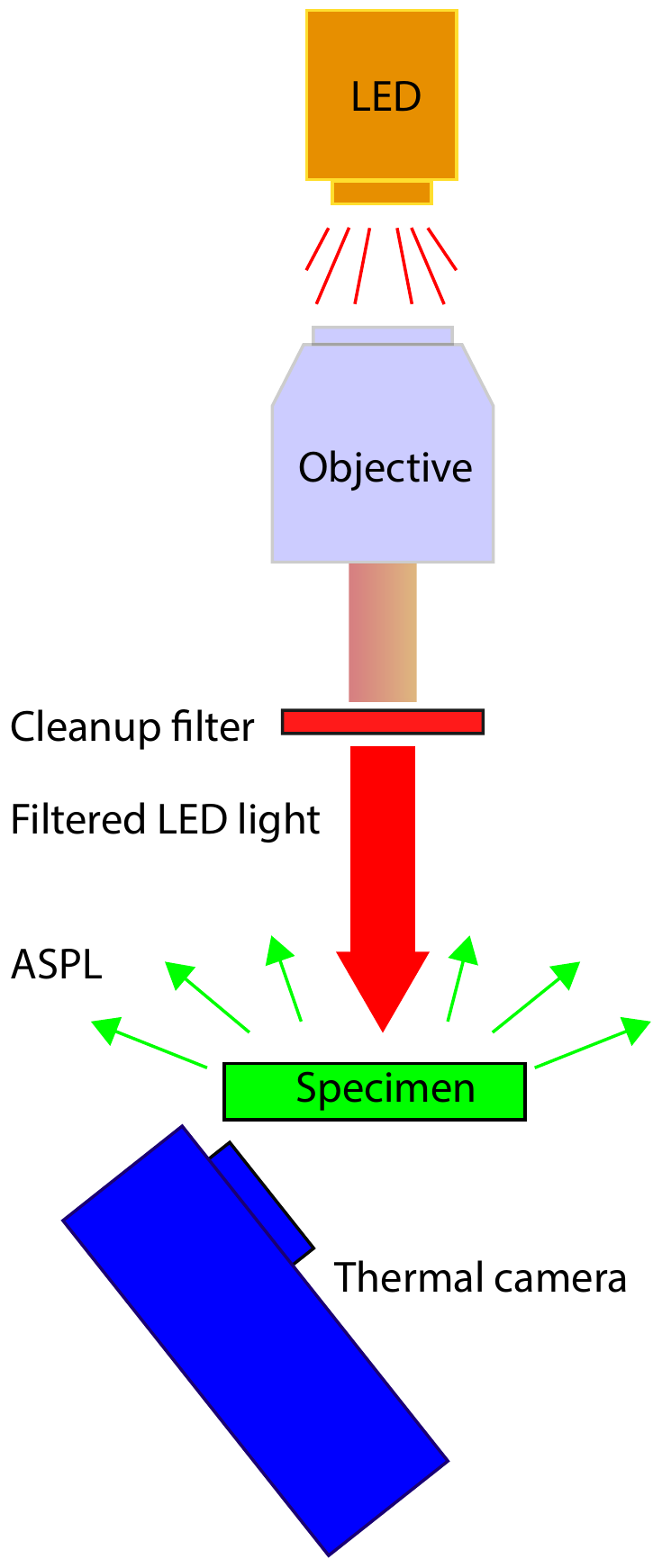}
    \caption{Schematic of proposed experiment to observe photoluminescent cooling with a light emitting diode (LED) as the photon source.}
    \label{fig:exp_proposal}
    \hrulefill
\end{figure}

\section{Discussion}
Our results show that using incoherent light sources like lamps, LEDs and filtered sunlight instead of lasers will not degrade optical cooling efficiencies so long as the source subtends a small solid angle (\textbf{Fig. 3}) and is sufficiently monochromatic so that the energy dispersion is less than the detuning. Using a low intensity beam with a larger diameter yields a greater cooling efficiency than a higher intensity beam for the same cooling power. Our predicted dependence of cooling efficiency with detuning is consistent with experimentally-reported values. \cite{seletskiy2010laser} 

The schematic of a proposal to experimentally observe optical cooling with an incoherent light source is shown in \textbf{Figure \ref{fig:exp_proposal}}. If the linewidth of the LED is greater than $\Delta E$, then a band pass filter must be used to achieve the required linewidth. Small temperature changes in the specimen can be observed with thermal imaging cameras, which have been used to measure the temperature of optically cooled samples. \cite{topper2021laser}

The possibility of using incoherent sources like lamps, LEDs and filtered sunlight for optical refrigeration dramatically broadens the scope of its applications. Compared to mechanical and thermoelectric refrigeration, optical refrigeration is free from vibrations, is more reliable, and is compact. This has applications in cooling infrared cameras, gamma-ray spectrometers, ultra-stable laser cavities, electron cryo-microscopy and low-noise amplifiers, where reliable, low mass, and vibration-free cooling is necessary. \cite{seletskiy2016laser}

\begin{acknowledgments}
 S.G., M.K., and B.J. thank the U.S. National Science Foundation (DMR-1952841) for financial support. We thank Zhuoming Zhang, Shubin Zhang, Mark A. Caprio, Shannon Dulz, and Patrick Fasano for helpful discussions.
\end{acknowledgments}


%

\end{document}